\documentclass[useAMS]{mn2e}
\usepackage{graphicx}
\usepackage{epsfig}
\usepackage{amssymb}
\usepackage{lscape}
\usepackage{ulem}
\usepackage{txfonts}

\def\logms{$\log$ (M$_{\star}$/M$_{\odot}$)}
\def\aq{ALMaQUEST}
\def\sigh2{$\Sigma_{\rm H_2}$}
\def\sigsfr{$\Sigma_{\rm SFR}$}

\def\sigstar{$\Sigma_{\star}$}
\def\dsfr{$\Delta$SFR}
\def\dsigsfr{$\Delta \Sigma_{\rm SFR}$}
\def\dsfe{$\Delta$SFE}
\def\dfgas{$\Delta f_{\rm H_2}$}

\def\c2s{C\,{\sc ii}$^{\star}$}

\def\dsfr{$\Delta$SFR}

\def\fgas{$f_{\rm H_2}$}

\title[ALMaQUEST III - Star formation efficiency] {The ALMaQUEST Survey:  III. Scatter in the resolved star forming main sequence is primarily due to variations in star formation efficiency.}

\author[Ellison et al.] {Sara L. Ellison$^1$,  Mallory D. Thorp$^1$, Lihwai Lin$^2$, Hsi-An Pan$^2$,  Asa F. L. Bluck$^3$,
  \newauthor    Jillian M. Scudder$^4$, Hossen Teimoorinia$^{5,1}$, Sebastian F. S\'{a}nchez$^6$, Mark Sargent$^7$\\ 
$^1$ Department of Physics \& Astronomy, University of Victoria, Finnerty Road, Victoria, British Columbia, 
  V8P 1A1, Canada\\
  $^2$ Institute of Astronomy \& Astrophysics, Academia Sinica, Taipei 10617, Taiwan\\
  $^3$ Kavli Institute for Cosmology \& Cavendish Astrophysics, University of Cambridge, Madingley Road, Cambridge, CB3 0HA, UK\\
  $^4$ Department of Physics and Astronomy, Oberlin College, Oberlin, Ohio, OH 44074, USA\\
  $^5$ NRC Herzberg Astronomy and Astrophysics, 5071 West Saanich Road, Victoria, BC V9E 2E7, Canada\\
  $^6$ Instituto de Astronom\'{i}a, Universidad Nacional Autonoma de Mexico, A. P. 70-264, C.P. 04510, Mexico, 
  D.F., Mexico\\
  $^7$ Astronomy Centre, Department of Physics and Astronomy, University of Sussex, Brighton BN1 9QH, UK
}

\begin{document}

\maketitle

\begin{abstract}
  Using a sample of 11,478 spaxels in 34 galaxies with molecular gas, star formation and stellar maps taken from the ALMA-MaNGA QUEnching and STar formation (\aq) survey, we investigate the parameters that correlate with variations in star formation rates on kpc scales.  We use a combination of correlation statistics and an artificial neural network to quantify the parameters that drive both the absolute star formation rate surface density (\sigsfr), as well as its scatter around the resolved star forming main sequence (\dsigsfr).  We find that \sigsfr\ is primarily regulated by molecular gas surface density (\sigh2) with a secondary dependence on stellar mass surface density (\sigstar), as expected from an `extended Kennicutt-Schmidt relation'.  However, \dsigsfr\ is driven primarily by changes in star formation efficiency (SFE), with variations in gas fraction playing a secondary role.  Taken together, our results demonstrate that whilst the absolute rate of star formation is primarily set by the amount of molecular gas, the variation of star formation rate above and below the resolved star forming main sequence (on kpc scales) is primarily due to changes in SFE.
  
\end{abstract}


\section{Introduction}

It has been known for over a decade that a galaxy's total star formation rate (SFR) correlates with its total stellar mass (e.g. Brinchmann et al. 2004; Noeske et al. 2007; Salim et al. 2007), a relation that has become widely known as the star forming main sequence (SFMS).  Although tight, there is real scatter around the SFMS, as star formation in low $z$ galaxies is modulated around its equilibrium value by a range of mechanisms such as bars and mergers (e.g. Ellison et al. 2011; Scudder et al. 2012).  Whilst the potential triggers of galactic starbursts are relatively well understood, the physics of this transformation are less clear and it remains a matter of considerable debate whether galaxies above the SFMS are there due to higher molecular gas fractions (\fgas=\sigh2/\sigstar, e.g. Scoville et al. 2016; Lee et al. 2017), or enhanced star formation efficiencies (SFE=\sigsfr/\sigh2, e.g. Saintonge et al. 2011; Genzel et al. 2015; Silverman et al. 2015, 2018), or a combination of both (Saintonge et al. 2012, 2016; Lin et al. 2017; Piotrowska et al. 2019), depending on whether the galaxy is located on or above the star forming main sequence (Sargent et al. 2014; Elbaz et al. 2018).  

A more recent discovery, that has been established from large integral field unit (IFU) surveys of low redshift galaxies, is that the global SFMS is due to the ensemble effect of a `resolved' SFMS on kpc scales  (rSFMS, e.g.  S\'{a}nchez et al. 2013; Cano-D\'{i}az et al. 2016; Hsieh et al. 2017).  In turn, Lin et al. (2019) have shown that the rSFMS is likely not fundamental, but rather due to the combination of two other more base relationships, one which regulates the amount of molecular gas in the interstellar medium (i.e. the molecular gas main sequence: \sigstar\ -- \sigh2) and the other that describes the conversion of the molecular gas into stars (i.e. the Kennicutt-Schmidt relation: \sigh2\ -- \sigsfr).

Despite the strong correlation between \sigstar\ and \sigsfr, significant galaxy-to-galaxy variations (Gonzalez-Delgado et al. 2016; Hall et al. 2018; Vulcani et al. 2019) contribute to scatter in the rSFMS.  The resolved scaling relations established by IFU surveys, complemented by spatially matched CO observations, provide excellent baselines from which to investigate the physics behind this scatter and hence understand star formation regulation on kpc scales (e.g. Bolatto et al. 2017; Utomo et al. 2017). By defining enhancements in \sigsfr\ relative to \sigh2\ (i.e. a boost in SFE) and a \sigh2\ enhancement relative to \sigstar\ (i.e. a boost in \fgas), we can quantify whether deviations from the rSFMS are more closely related to changes in SFE or gas fraction.

Using kpc-scale data to study the regulation of star formation has a further important advantage over previous studies that use global quantities.  There is clear evidence that there can be significant internal variations in SFE within a given galaxy (e.g.  Leroy et al. 2013; Utomo et al. 2017; Tomicic et al. 2019), and that the regulation of star formation happens on local, not global scales (Dey et al. 2019; Bluck et al. 2019b).  Studies that investigate correlations in global data will average over these internal differences, blurring the underlying physical relationships.

In this paper, we use the ALMA-MaNGA QUEnching and STar formation (\aq) survey (Lin et al. in prep), containing 46 galaxies with kpc-scale maps of molecular gas, stellar mass and SFR.  The \aq\ sample includes galaxies located both on (Lin et al. 2019), above (Ellison et al. 2019) and below (Lin et al. 2017; Pan et al. in prep) the global star forming main sequence.  This broad range of global properties translates to a broad range of local \sigsfr\ values, making it an ideal sample for investigating the mechanisms that regulate star formation.  

\section{Data}\label{data_sec}

The \aq\ survey contains 46 galaxies with 10 $<$ \logms\ $<$ 11.5 selected from the Mapping Nearby Galaxies at Apache Point Observatory (MaNGA; Bundy et al. 2015) survey and followed up with Atacama Large Millimeter Array (ALMA) CO (1-0) observations that are matched in spatial resolution (FWHM $\sim$ 2.5 arcsec) and mapped to the same grids of 0.5 arcsec pixels (0.3 kpc at the median redshift of the sample, $z$=0.03).  A full description of the ALMA observations, data reduction and molecular gas maps, which typically extend to 1-2 R/R$_e$, are presented in Lin et al. (in prep).  CO luminosities are converted to molecular gas surface densities (\sigh2) via a constant conversion factor  $\alpha_{CO}$ = 4.3 M$_{\odot}$ pc$^{-2}$ (K km s$^{-1}$)$^{-1}$ (an assumption that we will demonstrate does not affect our results), with a requirement that the S/N of the CO line intensity exceeds 3.  

Stellar mass surface densities (\sigstar) are taken from the public MaNGA Data Release 15 (DR15) data cubes, as processed by PIPE3D (S\'{a}nchez et al. 2016), with a minimum log \sigstar\ $>$ 7 M$_{\odot}$ kpc$^{-2}$ imposed (below which the rSFMS is poorly defined in our sample).  We compute de-projected galactocentric radii in units of the effective radius ($R_e$), which typically extend to 1.5-2 R/R$_e$, using the Sersic fit axial ratio from the NASA Sloan Atlas.  All radii quoted in this paper adopt these de-projected values.  Emission line fluxes are taken from the PIPE3D data cubes and corrected for internal extinction by assuming an intrinsic H$\alpha$/H$\beta$=2.85 and a Milky Way extinction curve (Cardelli, Clayton \& Mathis 1989).  Only spaxels classified as star-forming according to the diagnostic criterion of Kauffmann et al. (2003), with S/N $>$3 in the requisite emission lines and H$\alpha$ equivalent width $>$ 6 \AA\ are used for the current study.  Star formation rate surface densities (\sigsfr) are computed from H$\alpha$ luminosities, a technique that has been shown to reproduce the UV and IR SFRs well in IFU data (Catalan-Torrecilla et al. 2015).  Gas-phase metallicities (O/H) are determined using the O3N2 calibration of Marino et al. (2013).  Finally, we require the galaxy inclination to be less than 70 degrees ($b/a >$0.34), which removes 12 galaxies from the full \aq\ sample.   All surface densities in this paper are inclination corrected.  In total, 11,478 spaxels in 34 galaxies fulfill all of the above criteria. 

\begin{figure}
	\includegraphics[width=8cm]{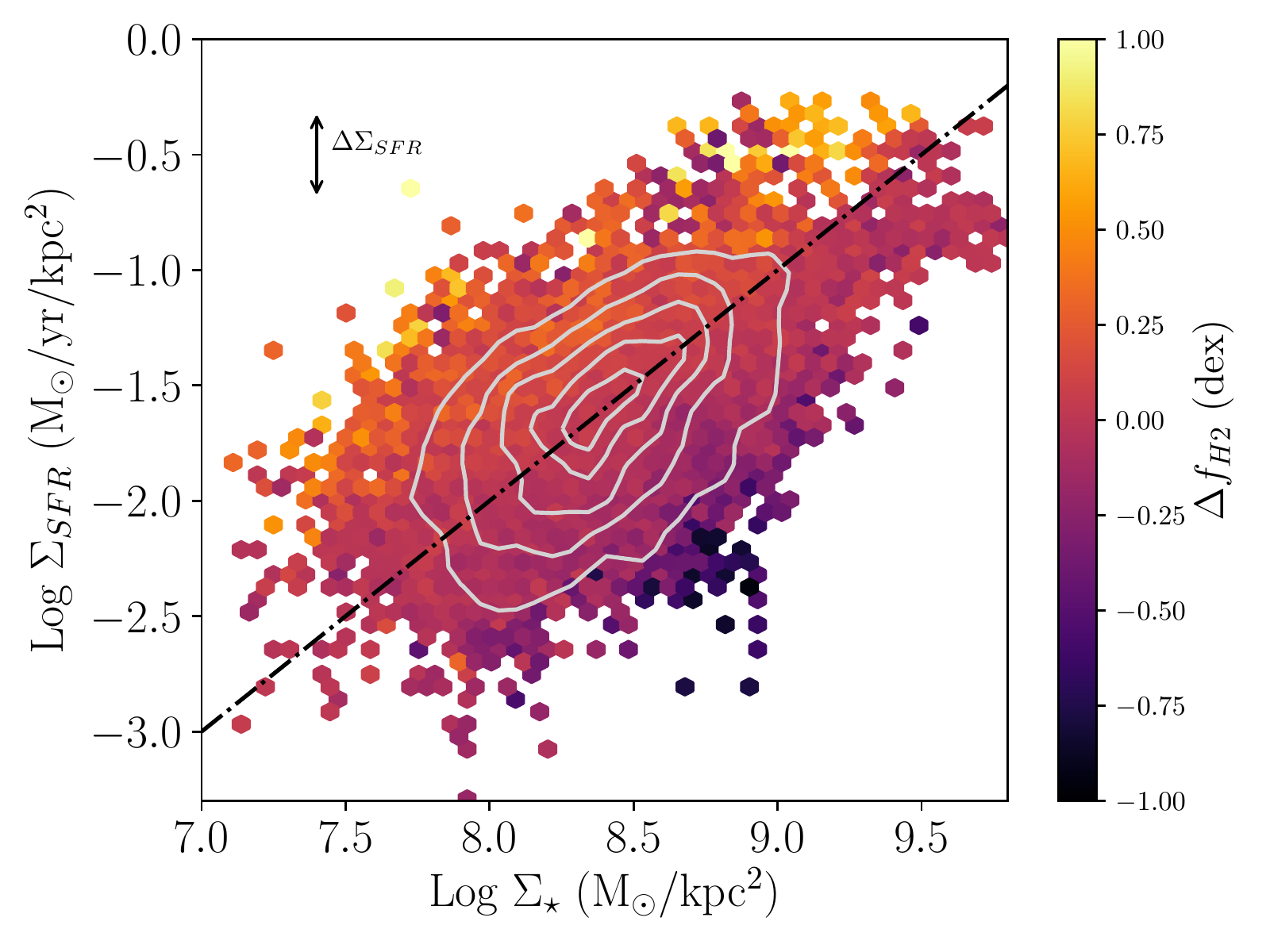}
	\includegraphics[width=8cm]{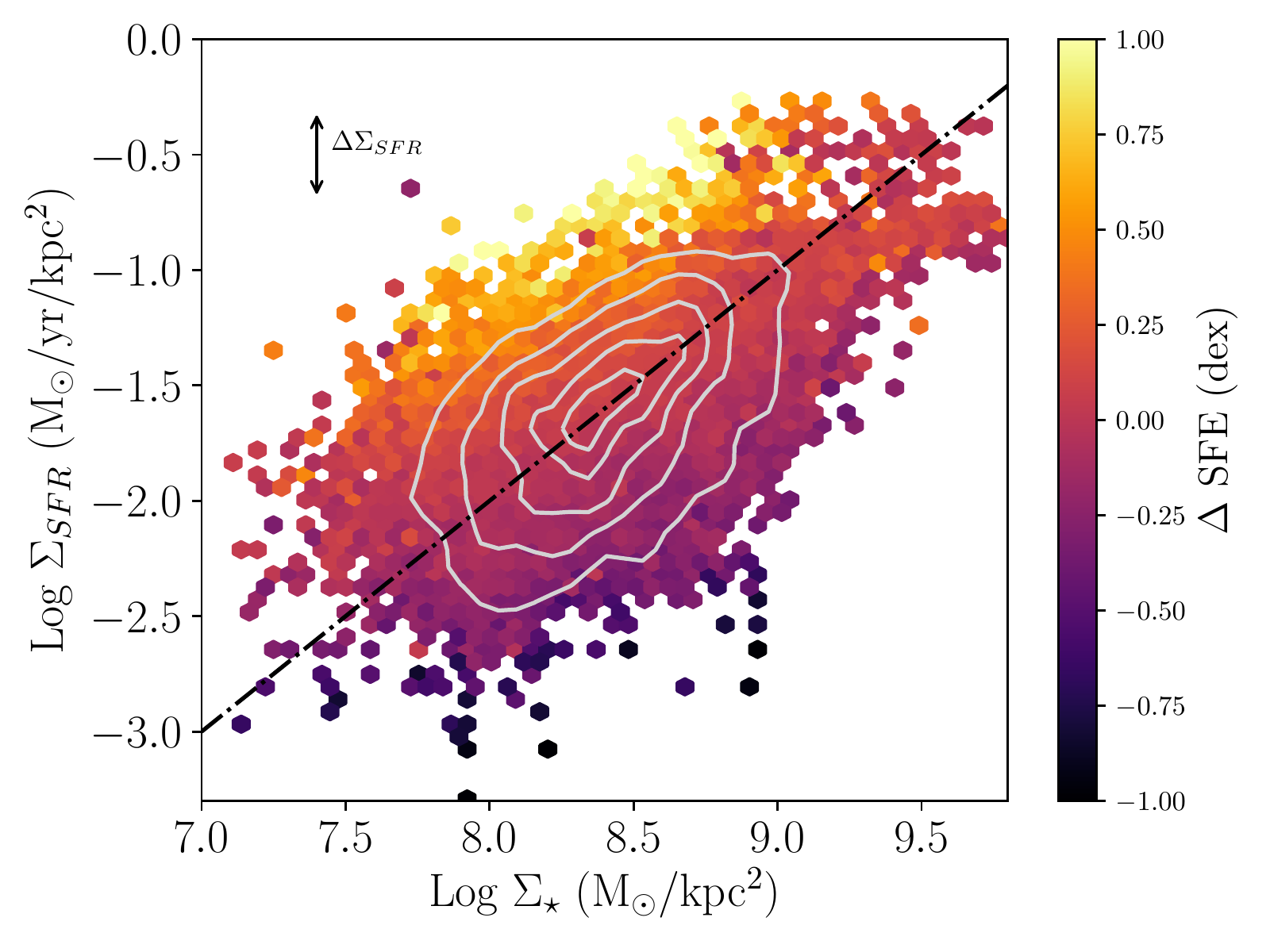}
        \caption{The binned resolved star forming main sequence for 11,478 spaxels in the \aq\ sample, where each grid cell is colour-coded by the mean \dfgas\ (top panel) or mean \dsfe\ (lower panel) of spaxels in that cell. For reference (i.e. not a fit to the data) the dot-dashed line shows a constant specific SFR=$10^{-10}$ yr$^{-1}$. Grey contours indicate the underlying spaxel density. }
    \label{ms_resolved_fig}
\end{figure}

In order to quantify how discrepant a given spaxel is from normal scaling relations, we follow Ellison et al. (2018, 2019) and compute a series of offsets from a set of control spaxels:

\begin{itemize}

\item \dsigsfr\ is the offset of a given spaxel's \sigsfr\ from the rSFMS of all ($\sim$ 1.2 million) star forming spaxels in the MaNGA DR15 at fixed \sigstar\ (matched within 0.1 dex).  \dsigsfr\ is therefore a measure of the `starburstiness' of a given spaxel, and encapsulates the scatter around the rSFMS. \\

\item \dsfe\ is the offset of a given spaxel's \sigsfr\ from the resolved Kennicutt-Schmidt relation relative to a set of control spaxels with normal SFRs ($-0.5 <$ \dsigsfr\ $<0.5$) that are matched in \sigh2\ (within 0.1 dex), $R/R_e$ (within 0.1 $R/R_e$), O/H (within 0.1 dex) and total stellar mass (within 0.1 dex).  Matching in O/H and $R/R_e$ mitigates metallicity and radius dependent $\alpha_{CO}$ variations (e.g. Sandstrom et al. 2013). \dsfe\ is therefore a measure of how efficient a given spaxel is at converting gas into stars, relative to the norm. \\

\item \dfgas\ is the offset of a given spaxel's \sigh2\ from the molecular gas main sequence (\sigstar\ -- \sigh2, e.g. Wong et al. 2013; Lin et al. 2019) relative to a set of control spaxels with normal SFRs ($-0.5 <$ \dsigsfr\ $<0.5$) that are matched in \sigstar\ (within 0.1 dex), $R/R_e$ (within 0.1 $R/R_e$), O/H (with 0.1 dex) and total stellar mass (within 0.1 dex). \dfgas\ is therefore a measure of how much extra gas, per unit stellar mass, is present in a given spaxel. \\
  
\end{itemize}

\section{Results}\label{results_sec}

In Fig. \ref{ms_resolved_fig} we show the rSFMS for the 11,478 star-forming spaxels (0.5 arcsec in size, sampling the 2.5 arcsec resolution) in the \aq\ sample, colour coded by \dfgas\ (top panel) and \dsfe\ (lower panel).  The relatively large scatter in this figure is a direct result of our sample selection, which contains a high fraction of both greeen valley and starburst galaxies.  Whilst this large dynamic range is advantageous for the study presented here (since our goal is to study the source of the scatter), it is not a representative (e.g. in mass or volume) sample, hence we do not report a fit to the rSFMS for comparison to other studies.  Fig. \ref{ms_resolved_fig} shows that the scatter in the rSFMS is clearly correlated with variations in a spaxel's gas fraction and SFE, in qualitative agreement with low redshift studies of global gas properties (e.g. Saintonge et al. 2012).       

\begin{figure}
	\includegraphics[width=8cm]{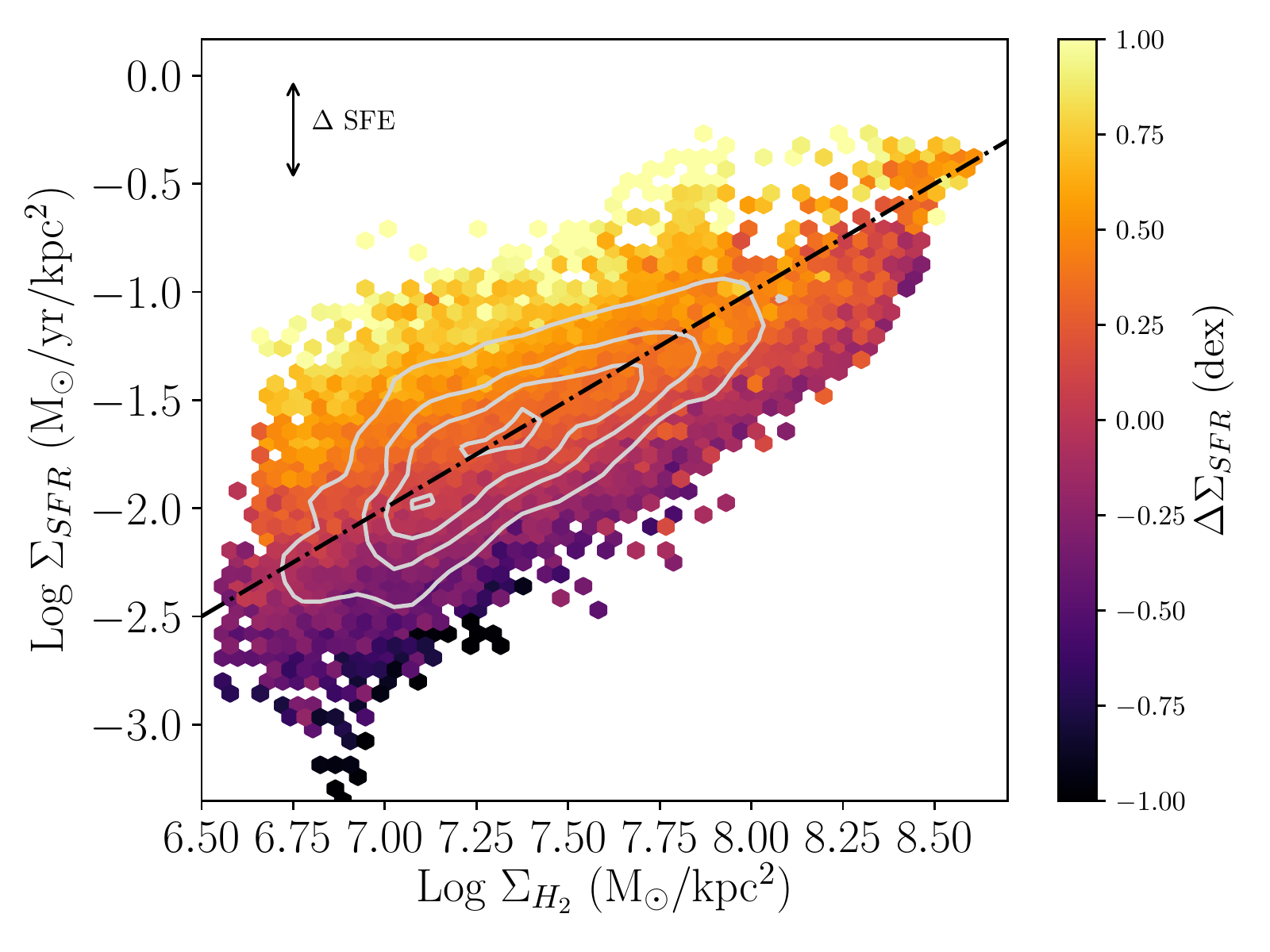}
	\includegraphics[width=8cm]{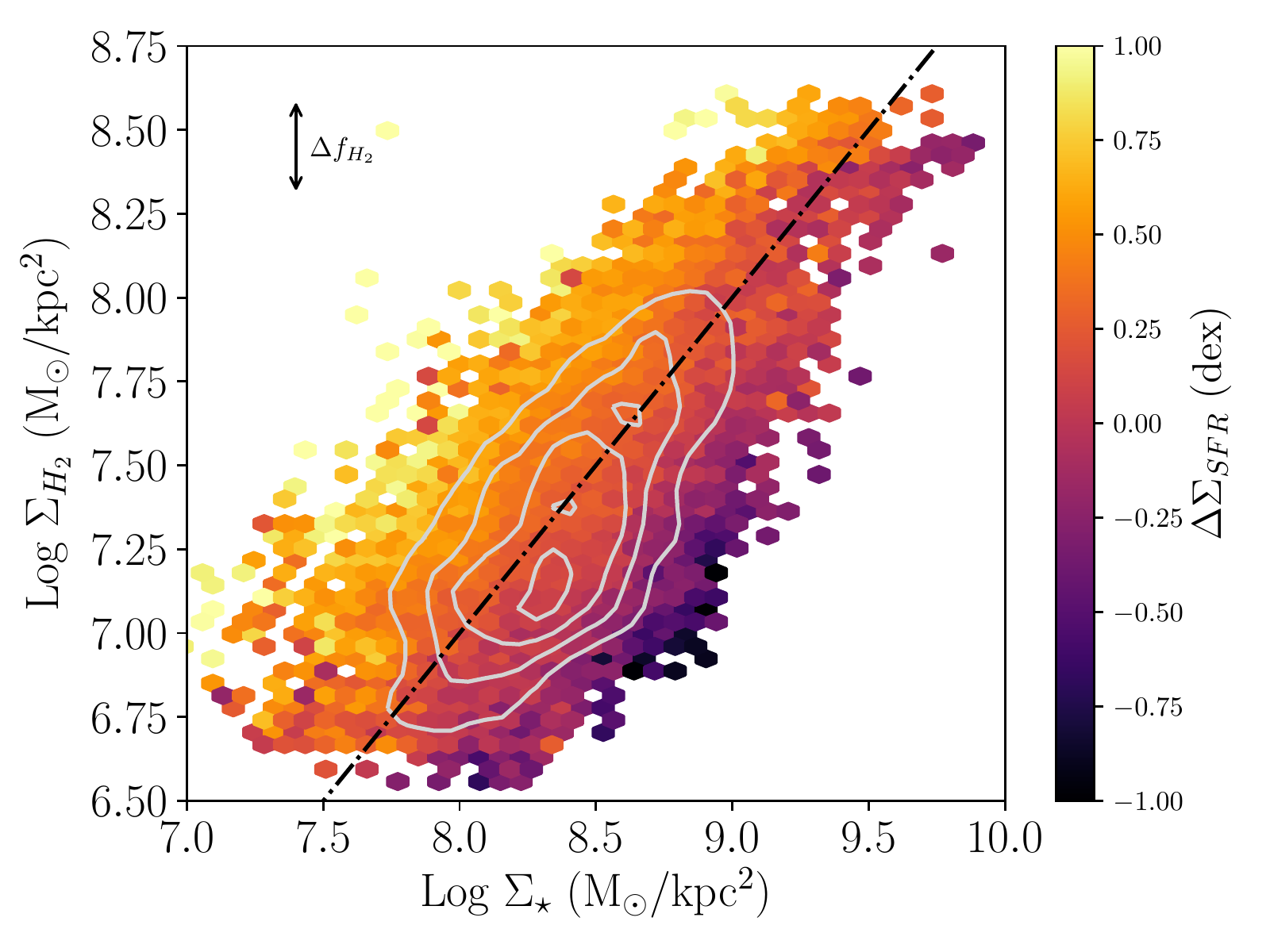}
        \caption{Upper panel: The binned resolved Kennicutt-Schmidt (\sigh2 -- \sigsfr) relation for 11,478 spaxels in the \aq\ sample, where each grid cell is colour coded by the mean \dsigsfr\ in that cell.  For reference (i.e. not a fit to the data) the dot-dashed line shows a constant depletion time of 1 Gyr.  Lower panel:  The binned resolved molecular gas main sequence (\sigstar\ -- \sigh2), colour coded by the  mean \dsigsfr\ in each grid cell.  For reference (i.e. not a fit to the data) the dot-dashed line shows a constant log \fgas=$-1$.  Grey contours indicate the underlying spaxel density.}
    \label{ks_resolved_fig}
\end{figure}

An alternative way to visualize the dependence of \dsigsfr\ on gas fraction and SFE is shown in Fig. \ref{ks_resolved_fig}, where we plot the distribution of  \sigh2\ vs. \sigsfr\ (upper panel) and \sigstar\ vs. \sigh2\ (lower panel), both colour-coded by \dsigsfr.  The upper panel of Fig \ref{ks_resolved_fig} shows the expected strong correlation between molecular gas and star formation (the Kennicutt-Schmidt relation).   To first order, a constant depletion time (the inverse of SFE) is a good representation of the data, consistent with studies of nearby star-forming disk galaxies (e.g. Bigiel et al. 2008, 2011; Leroy et al. 2008, 2013; Schruba et al. 2011).  However, the colour-coding reveals that a spaxel's SFE (i.e. its vertical position at fixed \sigh2) depends on its enhancement relative to the rSFMS.  The lower panel of  Fig. \ref{ks_resolved_fig} shows that the gas fraction of a given spaxel also depends on \dsigsfr, but the correlation is (visually) weaker than the relation between SFE and \dsigsfr\ (upper panel of Fig \ref{ks_resolved_fig}).  

In order to more rigourously quantify the relative importance of various parameters in affecting \sigsfr\ (the absolute surface density of star formation) and \dsigsfr\ (the scatter around the rSFMS), we perform two complementary ranking analyses.  First, we perform a simple correlation assessment in which we compute the Pearson correlation coefficient ($\rho$) between \sigsfr\ and \dsigsfr\ and a list of input variables.  Since Bluck et al. (2019b) have recently shown that global parameters have relatively little impact on kpc-scale star formation rates, we restrict our analysis to spaxel-based variables.  Specifically, we quantify $\rho$ for \sigstar, \sigh2, \fgas, SFE, \dsfe\ and \dfgas.  \dsfe\ and \dfgas\ are included in this list of variables as they capture subtly different information than the raw values of SFE and \fgas.  Fundamentally, these offset metrics quantify deviations from normal values, but here they have the additional power of accounting for dependences of $\alpha_{CO}$ on O/H and $R$.  

To complement the correlation analysis, we also apply a machine learning approach, which has the benefit of being able to recognize more complex patterns in the data than can be captured by $\rho$.    Specifically, we perform a non-linear regression with an artificial neural network (ANN), utilising MLPregresser from the scikit-learn package (Pedregosa et al. 2011), adopting an identical architecture as in Bluck et al. (2019b). To test the stability of the network, we train on five randomly selected input samples (containing 50\% of the data) and validate on the remainder of the sample in each case. We quantify the performance of the network with each individual parameter via the improvement over random (IOR) statistic (defined in Bluck et al. 2019a). Briefly, the mean square error of the network-predicted values to the truth values are computed in each case, and the improvement in performance over a random variable is quantified. In this manner parameters may be ranked in terms of how effective they are at predicting either \sigsfr\ or \dsigsfr.  We use the same six parameters as in the correlation analysis, all of which yield a highly stable result, with a typical variance of only $\sim$ 1\%.

\begin{figure}
  \includegraphics[width=8cm,trim={0 1.5cm 0 1.5cm}, clip]{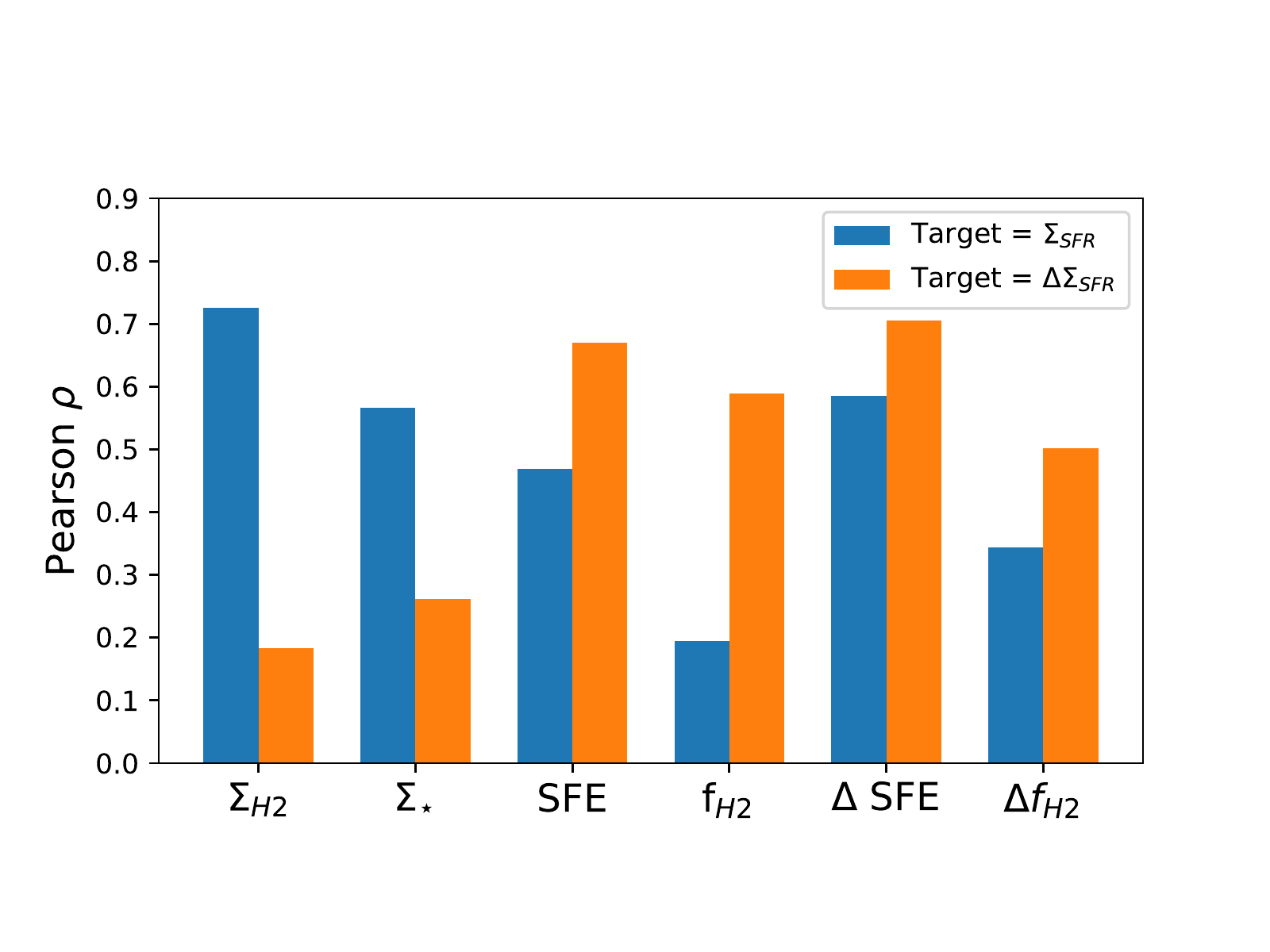}
	\includegraphics[width=8cm,trim={0 1.5cm 0 1.5cm},clip ]{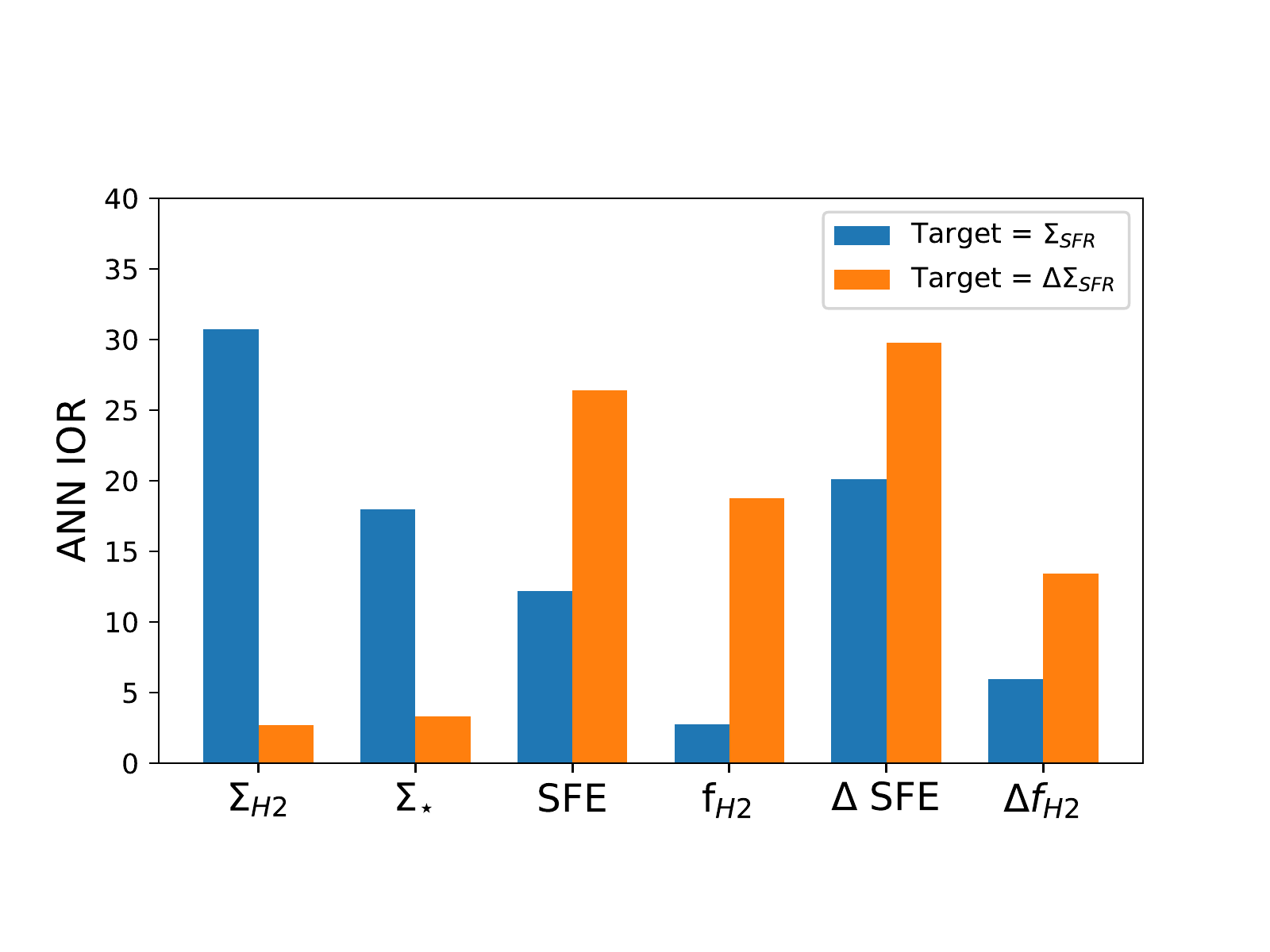}
        \caption{Performance of six input variables in estimating either \sigsfr\ (blue columns) or \dsigsfr\ (orange columns) for the correlation (upper panel) and ANN (lower panel) analyses.    }
    \label{bar_fig}
\end{figure}

Fig. \ref{bar_fig} shows the results from the correlation (upper panel) and ANN (lower panel) analyses of the six input variables in predicting either \sigsfr\ (blue columns) or \dsigsfr\ (orange columns) as the target variable, where a higher number ($\rho$ or IOR) indicates that a given variable is better at predicting the target variable.  Both analyses find that the single most important variable in predicting a spaxel's \sigsfr\ (of those we have tested) is \sigh2, i.e. the Kennicutt-Schmidt relation.  Variations in SFE (i.e \dsfe) and \sigstar\ play equally important secondary roles.  The former of these two variables captures the scatter around the Kennicutt-Schmidt relation, and accounts for variations in O/H and radius.  The equally important role of \sigstar\ is consistent with expectations from an `extended Kennicutt-Schmidt' law in which \sigsfr\ $\propto$ \sigh2$^n$\sigstar$^m$, where observations find $m<n$ (e.g. Shi et al. 2011).

\begin{figure}
	\includegraphics[width=8cm]{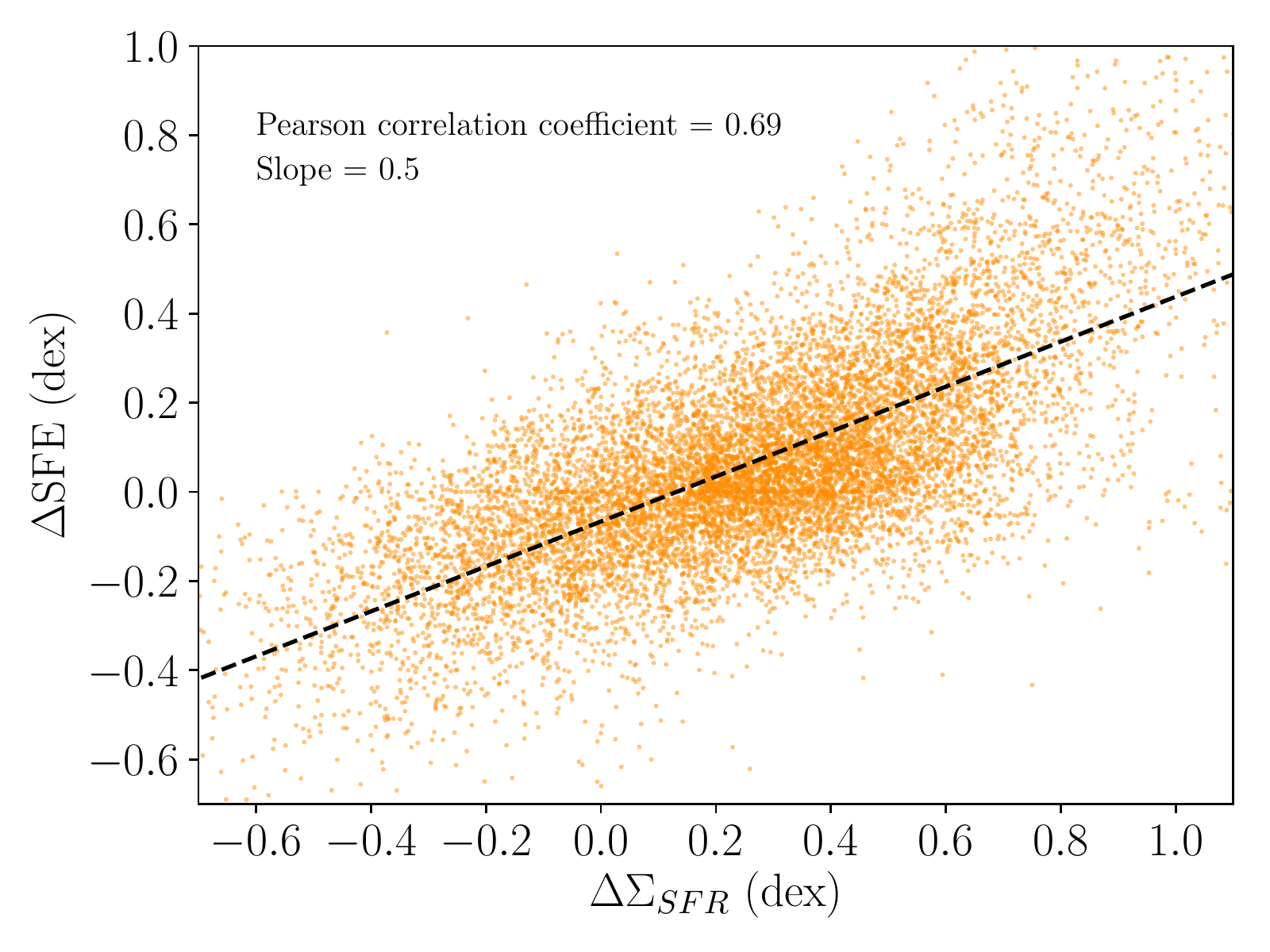}
        \caption{Correlation between \dsigsfr\ and \dsfe.   }
    \label{corr_fig}
\end{figure}

A subtly different question, and the main objective of the current paper, is assessed by the orange bars in Fig. \ref{bar_fig}, in which the target variable is now set not to be \sigsfr, but its offset from the rSFMS, \dsigsfr.  The results from this test tell us which factors correlate with a spaxel's SFR deviation from the norm, causing it to be either enhanced, or suppressed, in its star formation.  Again, the results of the correlation analysis are in good qualitative agreement with the ranks obtained from the more complex ANN approach.  Both analyses find that, in contrast to predicting the absolute \sigsfr, \sigh2\ plays a minor role in predicting \dsigsfr.  Instead, \dsfe\ is found to be the most important parameter for driving \sigsfr\ away from the rSFMS, with a secondary dependence on \fgas, indicating that the scatter around the rSFMS is primarily driven by by changes in the SFE.   The correlation between \dsigsfr\ and \dsfe\ is shown explicitly in Fig. \ref{corr_fig}.  We repeated our ANN regression analysis separating spaxels into those on ($-0.5 <$ \dsigsfr\ $<0.5$) and above (\dsigsfr $>$ 0.5) the rSFMS and found that \dsfe\ remains the most important parameter for predicting  \dsigsfr\ in both cases.

\section{Summary and Discussion}\label{summary_sec}

The key result of the work presented here is that, on kpc-scales, the enhancement of star formation away from its equilibrium value (as defined by the rSFMS) is primarily driven by changes in SFE, with a secondary, weaker, dependence on gas fraction.  This conclusion is based on (1) assessment of Pearson correlation coefficients and (2)  an artificial neural network analysis.  Furthermore, we find that \dsfe\ drives \dsigsfr\ both within and above the rSFMS.

Throughout this work we have adopted a constant $\alpha_{CO}$ conversion factor to determine \sigh2.   In practice, $\alpha_{CO}$ is likely to be lower for spaxels with enhanced \dsigsfr, leading to a decrease in \sigh2.  In turn, a decrease in \sigh2\ would lead to an increased SFE (and \dsfe) and a decrease in \fgas\ (and \dfgas).   Our correlation analysis already identifies a strong positive correlation between \dsigsfr\ and \dsfe; increasing \dsfe\ would further strengthen that correlation, e.g. steepening the correlation in Fig. \ref{corr_fig}.  Likewise, in the ANN analysis we see that \dsigsfr\ is better predicted by \dsfe\ than absolute SFE, probably because O/H and $R$ (which both affect $\alpha_{CO}$) are controlled for in the former and hence \dsfe\ better captures variations in the conversion factor.  On the other hand, the reduction in \dfgas\ that would result from a lower $\alpha_{CO}$ would reduce the strength of the positive correlation with \dsigsfr\ and indeed we see that \dfgas\ performs less well than \fgas\ in the ANN analysis.  Our conclusion that \dsigsfr\ is more correlated with \dsfe\ than \dfgas\ would therefore only be strengthened if we explicitly used a variable conversion factor. 

Previous studies of \textit{global} quantities have also concluded that there is a correlation between the position of a galaxy relative to the main sequence (\dsfr) and its total SFE (Genzel et al. 2015; Silverman et al. 2015, 2018; Tacconi et al. 2018).  Our results agree with this primary dependence, but show a secondary depedence on gas fraction (e.g. Saintonge 2012, 2016).  However, global studies average over the considerable internal variations known to exist within a given galaxy (e.g. Tomicic et al. 2019; Vulcani et al. 2019), which will induce scatter into relations and obscure the underlying scale which governs the physics.  Our sample also has the significant advantage of being drawn from a single dataset with continuous selection function (rather than focused on high SFR or CO-bright samples), single redshift epoch, homogeneous data reduction and consistent assumptions on, for example, definition of main sequence offset and treatment of $\alpha_{CO}$.  Any correlations in our data are therefore unlikely to be due to combining data from different sources, different redshifts (where different physical mechanisms may be at play) with very different properties and analysis methods.

Whereas \dsigsfr\ is set primarily by changes in SFE, our ANN analysis shows that \sigsfr\ is governed mainly by \sigh2, consistent with expectations from the Kennicutt-Schmidt relation, with a secondary dependence on \sigstar\ (as well as a on \dsfe\ which is essentially the scatter around the Kennicutt-Schmidt relation).  Two other works have also recently used complementary machine learning methods to quantify the primary driver of \sigsfr.  First, Bluck et al. (2019b) have applied the same ANN approach used herein to MaNGA data and find several conclusions in common with us: that \sigstar\ is the most important parameter in determining \sigsfr, and that $R$ and O/H play a negligible role.  However, Bluck et al. (2019b) do not have direct measurements of gas surface densities amongst their available parameters.  Our results are therefore consistent with those of Bluck et al. (2019b) in that of the non-gas parameters in our sample, we also find \sigstar\ as the highest ranked parameter in predicting \sigsfr.

Second, Dey et al. (2019) combine IFU data from the Calar Alto Legacy Intergral Field Area (CALIFA, S\'{a}nchez et al. 2012) survey, with interferometric observations of CO, to achieve a dataset that is similar in its range of physical variables to the \aq\ sample studied herein.  Even with the inclusion of molecular gas maps, Dey et al. (2019) agree with the conclusion of Bluck et al. (2019b) on the primary importance of \sigstar\ in determining \sigsfr, ranking it above \sigh2.  Dey et al. (2019)'s result therefore contrasts with the conclusions of our study, and imply that, at kpc-scales, the rSFMS is more fundamental than the Kennicutt-Schmidt law.  The results of Dey et al. (2019) are surprising in light of the well-established link between gas and star formation and are also in conflict with Lin et al. (2019), who conclude that the rSFMS is largely due to the more fundamental \sigstar -- \sigh2\ and \sigh2 -- \sigsfr\ relations.

Although we have concluded that variations in SFE are responsible for variations in \sigsfr\ on kpc-scales, a potentially important omission in our study is the role of dense molecular gas.   Gao \& Solomon (2004) showed that a linear relationship exists between the global abundance of dense gas (as traced by HCN) in galaxies and their SFR, implying a constant SFE in the dense gas (see also Wu et al. 2005; Lada et al. 2010).  These works have supported a so-called `threshold model' of star formation, whereby the abundance of dense gas is a direct indicator of how much star formation takes place.  However, more recent work, that has mapped dense gas on kpc-scales, has found that there is real scatter in the relation between dense gas and SFR within a given galaxy (e.g. Usero et al. 2015; Bigiel et al. 2016).   Contrary to a threshold star formation model, these works demonstrate that the star formation efficiency of dense gas is very sensitive to local parameters, such as \sigstar\ (Gallagher et al. 2018; Jimenez-Donaire et al. 2019).  Including measurements of the dense gas surface density in an ANN analysis is likely to yield further insights into the primary regulators of star formation on kpc scales.

\section*{Acknowledgements}

We gratefully acknowledge grant MOST 107-2119-M-001-024 for funding travel to ASIAA (SLE and MDT) and ERC Advanced Grant 695671 `QUENCH' (AFLB). SLE thanks Erik Rosolowsky and Brenda Matthews for discussions and ALMA advice.  This paper makes use of the following ALMA data: ADS/JAO.ALMA\#2015.1.01225.S, ADS/JAO.ALMA\#2017.1.01093.S, ADS/JAO.ALMA\#2018.1.00558.S, ADS/JAO.ALMA\#2018.1.00541.S.   ALMA is a partnership of ESO (representing its member states), NSF (USA) and NINS (Japan), together with NRC (Canada), MOST and ASIAA (Taiwan), and KASI (Republic of Korea), in cooperation with the Republic of Chile. The Joint ALMA Observatory is operated by ESO, AUI/NRAO and NAOJ.  The National Radio Astronomy Observatory is a facility of the National Science Foundation operated under cooperative agreement by Associated Universities, Inc.


\end{document}